\documentclass[12pt]{iopart}
\usepackage{iopams}
\usepackage{graphicx}

\begin{document}

\title{Applications of the Gauss-Bonnet theorem to gravitational lensing}
\author{ G W Gibbons$^1$ and M C Werner$^2$} 
\address{$^1$ Department of Applied Mathematics and Theoretical Physics, University of Cambridge, Wilberforce Road, Cambridge CB3~0WA, UK}
\ead{G.W.Gibbons@damtp.cam.ac.uk}
\address{$^2$ Institute of Astronomy, University of Cambridge, Madingley Road, Cambridge CB3~0HA, UK}
\ead{mcw36@ast.cam.ac.uk}

\begin{abstract}
In this geometrical approach to gravitational lensing theory, we apply the Gauss-Bonnet theorem to the optical metric of a lens, modelled as a static, spherically symmetric, perfect non-relativistic fluid, in the weak deflection limit. We find that the focusing of the light rays emerges here as a topological effect, and we introduce a new method to calculate the deflection angle from the Gaussian curvature of the optical metric. As examples, the Schwarzschild lens, the Plummer sphere and the singular isothermal sphere are discussed within this framework.
\end{abstract}


\submitto{\CQG}
\maketitle

\section{Introduction}
The deflection of light by gravitational fields has been studied with great interest in astrophysics as well as in theoretical physics. Fundamental properties such as Fermat's principle for Lorentzian manifolds, conditions on image multiplicity and caustics in spacetime have been discussed in a fully relativistic setting (e.g., see \cite{perlick} and references therein). In the astrophysical context, however, an impulse approximation with piecewise straight light rays in flat space has proven useful since deflection angles on cosmological scales are very small (for a comprehensive introduction, see e.g. \cite{schneider} or \cite{straumann} pp 272--95 and references therein). Despite their different premisses, both treatments have yielded mathematically interesting, general properties which depend on topology. In particular, image counting theorems like the odd number theorem have been established with different versions of Morse theory both in the spacetime lensing and impulse approximation frameworks \cite{mckenzie}.

In this article, we would like to present another approach to gravitational lensing theory which emphasizes global properties. Specifically, we consider the astrophysically relevant weak deflection limit not in the impulse approximation but treat light rays as spatial geodesics of the optical metric, and use the Gauss-Bonnet theorem. This approach has previously been applied to lensing by cosmic strings \cite{gibbons}, and we extend it here to static, spherically symmetric bodies of a perfect fluid as simple models for galaxies acting as gravitational lenses. It turns out that the focusing of light rays is, from this point of view, essentially a topological effect. Hence we find, rather surprisingly, that the deflection angle can be calculated by integrating the Gaussian curvature of the optical metric outwards from the light ray, in contrast to the usual description in terms of the mass enclosed within the impact parameter of the light ray. To illustrate, we discuss how this works for three well-known models, namely the Schwarzschild lens, the Plummer sphere and the singular isothermal sphere.

The structure of this article is therefore as follows. In section 2 we give a brief review of the Gauss-Bonnet theorem and introduce two constructions which will be used to investigate the lensing geometry. The optical metric and its Gaussian curvature for static, spherically symmetric systems of a perfect fluid is discussed in section 3, followed by the application to the three lens models mentioned above in section 4.

With regard to conventions, we use metric signature $(-,+,+,+)$, Latin and Greek indices for space and spacetime coordinates, respectively, and set the speed of light $c=1$. $G$ denotes the gravitational constant as usual.

\section{Gauss-Bonnet theorem and lensing geometry}
\begin{figure}
\centering
\includegraphics[width=0.5\columnwidth, height=0.35\columnwidth]{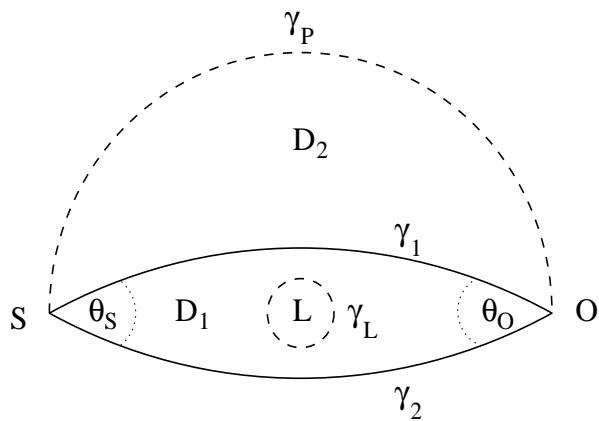}
\caption{Weak deflection lensing geometry. Two geodesics $\gamma_1$ and $\gamma_2$ from the source $S$ to the observer $O$ are deflected by a lens with centre at $L$. $D_1$ and $D_2$ are two domains with boundary curves $\gamma_L$ and $\gamma_P$ as discussed in the text.}
\label{fig:gaussbonnet}
\end{figure}

The Gauss-Bonnet theorem connects the intrinsic differential geometry of a surface with its topology, and this will be the main tool in our exposition. Let the domain $(D,\chi,g)$ be a subset of a compact, oriented surface, with Euler characteristic $\chi$ and a Riemannian metric $g$ giving rise to a Gaussian curvature $K$. Furthermore, let $\partial D:\{t\}\rightarrow D$ be its piecewise smooth boundary with geodesic curvature $\kappa$, where defined, and exterior angle $\alpha_i$ at the $i$th vertex, traversed in the positive sense. Then the local and global versions of the Gauss-Bonnet theorem (see, e.g., \cite{klingenberg} pp 139, 143) can be combined to give
\begin{equation}
\int\int_D K \rmd S + \int_{\partial D} \kappa \rmd t +\sum_i \alpha_i = 2\pi \chi(D).
\label{gb}
\end{equation}
Next, consider a smooth curve $\gamma:\{t\}\rightarrow D$ of unit speed such that $g(\dot{\gamma},\dot{\gamma})=1$, and let $\ddot{\gamma}$ be the unit acceleration vector. This vector, then, is perpendicular to $\dot{\gamma}$ and hence spans a Frenet frame together with $\dot{\gamma}$. The geodesic curvature of $\gamma$ is therefore given by (see, e.g., \cite{klingenberg} p 138)
\begin{equation}
\kappa=g\left(\nabla_{\dot{\gamma}}\dot{\gamma},\ddot{\gamma}\right),
\label{kappa}
\end{equation} 
which clearly vanishes if, and only if, $\gamma$ is a geodesic. We will now describe two specific setups for the domain, $D_1$ and $D_2$, which will represent the weak deflection lensing geometries discussed in section 4, and are shown in figure \ref{fig:gaussbonnet}.

Firstly, let $D_1$ be bounded by two geodesics $\gamma_1,\gamma_2$ intersecting in two vertices, the first being the source $S$, and the second the observer $O$. Hence $\kappa(\gamma_1)=\kappa(\gamma_2)=0$ except at the source and observer where the interior angles are $\theta_S=\pi-\alpha_S$ and $\theta_O=\pi-\alpha_O$, respectively. $D_1$ contains the lens centre $L$, and both $S$ and $O$ are assumed to be very distant from $L$ such that $\theta_S$ and $\theta_O$ are small and positive. If the lens centre is non-singular, then $D_1$ is simply connected so that $\chi(D_1)=1$. On the other hand, if $L$ is singular, then $\chi(D_1)=0$ and there is a contribution to $\partial D_1$ given by the boundary curve $\gamma_L$ enclosing $L$. Hence the Gauss-Bonnet theorem (\ref{gb}) yields
\begin{equation}
\theta_S+\theta_O=\int\int_{D_1} K \rmd S
\label{gb1}
\end{equation}
in the non-singular case, and
\begin{equation}
\theta_S+\theta_O=2\pi+\oint_{\gamma_L}\kappa dt +\int\int_{D_1} K \rmd S
\label{gb2}
\end{equation}
in the singular case.

Secondly, let $D_2$ be a simply connected domain such that $L\notin D_2$ and $D_1 \cap D_2=\gamma_1$. Furthermore, $D_2$ be asymptotically flat such that $S$ and $O$ are in this regime. Thus $\partial D_2$ consists of a geodesic $\gamma_1$ from $S$ to $O$, and the perimeter curve $\gamma_P$ of a circular sector centred on $L$ intersecting $\gamma_1$ in $S$ and $O$ at right angles. If $\phi$ is the angular coordinate centred on $L$, then $\gamma_p$ has angular range $\pi+\delta$ where $\delta$ is the small and positive, asymptotic deflection angle of $\gamma_1$. Therefore we can use (\ref{gb}) to write 
\begin{equation}
\int_0^{\pi+\delta}\kappa (\gamma_P) \frac{\rmd t}{\rmd \phi} \rmd \phi -\pi=-\int\int_{D_2} K \rmd S,
\label{gb3}
\end{equation}
again using $\kappa(\gamma_1)=0$. By construction, the calculation of the deflection angle according to (\ref{gb3}) is obviously independent of any singularity that may occur at $L$ in $D_1$.

\section{Optical metric and its Gaussian curvature}

We shall now consider null geodesics deflected by a static, spherically symmetric massive body consisting of a perfect fluid. In the present context, this is to be thought of as the stellar fluid of a galaxy acting as a gravitational lens. The corresponding line element written in Schwarzschild coordinates $x^\mu=(t,r,\vartheta,\phi)$ is given by (e.g., see \cite{straumann} p 375)
\begin{eqnarray}
\rmd s^2&=g_{\mu\nu}\rmd x^\mu \rmd x^\nu \nonumber \\
&= -\exp[2A(r)]\rmd t^2+\exp[2B(r)]\rmd r^2+r^2\left(\rmd \theta^2+\sin^2\vartheta \rmd \phi^2\right).
\label{metric}
\end{eqnarray}
By spherical symmetry, we can assume, without loss of generality, that the null geodesics with $\rmd s^2=0$ are in the equatorial plane $\vartheta=\pi/2$. All images are therefore collinear with the lens centre $L$ in this case. Spatial projection of the null geodesics yields the light rays, and these are geodesics of the optical metric $g^{\mathrm{opt}}_{mn}=g_{mn}/(-g_{00})$, by Fermat's principle. The geometry of the optical metric, also known as optical reference geometry, is useful for the study of gravitational and inertial forces in general relativity \cite{abram}. It is convenient to introduce a radial Regge-Wheeler tortoise coordinate $r^*$ defined by
\[
\rmd r^*=\rmd r \exp[B(r)-A(r)],
\]
so that the line element of the optical metric becomes
\begin{eqnarray}
\rmd t^2 &= g^{\mathrm{opt}}_{mn}\rmd x^m \rmd x^n=\exp[2(B-A)]\rmd r^2+\exp(-2A)r^2\rmd \phi^2 \nonumber \\
&= \rmd r^{*2}+f(r^*)^2\rmd \phi^2
\label{optmetric}
\end{eqnarray}
from (\ref{metric}), where $f(r^*)=\exp[-A(r)]r, \ r=r(r^*)$. It is clear, then, that the equatorial plane in the optical metric is a surface of revolution when it is embedded, where possible, in $\mathbb{R}^3$. Its (intrinsic) Gaussian curvature $K$ (e.g., see \cite{klingenberg} p 66) can be found from (\ref{optmetric}) and expressed in terms of the Schwarzschild radial coordinate $r$,
\begin{eqnarray}
K&=-\frac{1}{f(r^*)}\frac{\rmd^2 f}{\rmd r^{*2}}\nonumber \\
&=-\exp[2(A-B)]\left(\frac{\rmd A}{\rmd r}\frac{\rmd B}{\rmd r}-\frac{1}{r}\frac{\rmd A}{\rmd r}-\frac{1}{r}\frac{\rmd B}{\rmd r}-\frac{\rmd^2 A}{\rmd r^2}\right).
\label{gauss}
\end{eqnarray}
One can now specialize to the perfect fluid case. Then the energy-momentum tensor is $T^{\mu\nu}=\mathrm{diag}(\rho,p,p,p)$ in the local flat metric, where $\rho=\rho(r)$ denotes the density and $p=p(r)$ the pressure of the lens model. Now from Einstein's field equations (e.g., see \cite{straumann} pp 376--7),
\begin{equation}
\exp[-2B(r)]=1-\frac{2\mu(r)}{r} \qquad \mbox{where} \qquad \mu(r)=4\pi G \int_0^r \rho(r')r'^2 \rmd r'.
\label{b}
\end{equation}
The conservation of the energy-momentum tensor, together with the field equations, gives rise to the Tolman-Oppenheimer-Volkoff equation
\begin{equation}
\frac{\rmd p}{\rmd r}=-\frac{(\rho+p)\left(\mu+4\pi G r^3 p\right)}{r^2\left(1-\frac{2\mu}{r}\right)},
\label{tov}
\end{equation}
which expresses the hydrostatic equilibrium, and we also have
\begin{equation}
\frac{\rmd A}{\rmd r}=\frac{1}{1-\frac{2\mu}{r}}\left(\frac{\mu}{r^2}+4\pi G p r\right).
\label{a}
\end{equation}
Hence using (\ref{gauss}), the Gaussian curvature is
\begin{equation*}
\fl
K=-\frac{2\mu\exp[2(A-B)]}{r^3\left(1-\frac{2\mu}{r}\right)^2}\left\{1-\frac{3\mu}{2r}-\frac{4\pi Gr^3}{\mu}\left[\rho\left(1-\frac{2\mu}{r}\right)+p\left(1-\frac{3\mu}{r}-2\pi Gp r^2\right)\right]\right\}.
\end{equation*}
Since we are presently interested in weak deflections only, one can limit the discussion to a non-relativistic stellar fluid subject to the collisionless Boltzmann equation (e.g. \cite{binney}). But this means that the pressure may be neglected in (\ref{gauss}). To see this, recall that non-relativistic kinetic theory, for instance in the case of an isotropic velocity dispersion $\sigma^2$, implies that $p=\rho\sigma^2/3$ but $\sigma^2\ll1$. Now the area element of the equatorial plane in the optical metric is
\[
\rmd S=\sqrt{|\det g^{\mathrm{opt}}|}\rmd r \rmd \phi =\exp[B(r)-2A(r)]r \rmd r \rmd \phi
\]
from (\ref{optmetric}). Together with (\ref{b}), the integrand of the Gaussian curvature therefore reduces to
\begin{equation}
K\rmd S=-\frac{2\mu}{r^2\left(1-\frac{2\mu}{r}\right)^{3/2}}\left[1-\frac{3\mu}{2r}-\frac{4\pi Gr^3\rho}{\mu}\left(1-\frac{2\mu}{r}\right)\right]\rmd r \rmd \phi.
\label{gauss2}
\end{equation}
Notice that for realistic lens models, where $\rho\rightarrow \rho_0>0$ as $r\rightarrow 0$, and $\mu\rightarrow \mu_\infty<\infty$ as $r \rightarrow \infty$, the Gaussian curvature must change sign at some radius in the equatorial plane. For if $r\rightarrow \infty$, then $\rho$ has to decrease faster than $r^{-3}$ for $\mu$ to be asymptotically finite, so $K<0$ follows immediately from (\ref{gauss2}). Conversely, if $r\rightarrow 0$, then $\mu(r) \rightarrow 4\pi G\rho_0 r^3/3$ by (\ref{b}), so the term in the square bracket in (\ref{gauss2}) tends to $-2$, and the result follows. 

In the next section, we shall discuss three particular lens models as examples.

\section{Applications to lensing models}
\subsection{Schwarzschild lens}
The Schwarzschild lens can be characterized by $\mu(r)=\mathrm{const.}$ and $\rho(r)=0, \ r>0$ so that outside the event horizon at $r=2\mu$,
\begin{equation}
K\rmd S=-\frac{2\mu}{r^2\left(1-\frac{2\mu}{r}\right)^{3/2}}\left(1-\frac{3\mu}{2r}\right) < 0
\label{schwarz}
\end{equation}
from (\ref{gauss2}). The optical Schwarzschild metric has therefore negative Gaussian curvature, and an embedding diagram illustrating this can be found e.g. in \cite{abram}. Now, in light of this fact, it might appear surprising at first glance that light rays can be focused. For consider a geodesic $\gamma(t)$ separated from a neighbouring geodesic by a Jacobi field $Y(t)=y(t)\ddot{\gamma}$, then the geodesic deviation equation becomes (see, e.g., \cite{klingenberg} p 102)
\[
\frac{\rmd^2 y}{\rmd t^2}+K(\gamma)y=0,
\]
implying that the light rays should diverge. However, given a domain $D_1$ as described in section 2, the focusing of two light rays at the observer's vertex $O$ is made possible by the non-trival topology of $D_1$ such that $\chi(D_1)=0$ because of the event horizon. It is convenient to use the circular photon orbit at $r=3\mu$ as inner boundary curve $\gamma_L$ because its geodesic curvature vanishes. Then equation (\ref{gb2}) applies,
\[
\theta_S+\theta_O=2\pi+\int\int_{D_1} K \rmd S,
\]
which can clearly be fulfilled for small, positive $\theta_S,\theta_O$ despite the negative Gaussian curvature. The topological contribution is therefore essential to the focusing, and it is in this sense that gravitational lensing can be understood as a topological effect. In fact, its global nature is also borne out by equation (\ref{gb3}). Here, the deflection angle is calculated by integrating the Gaussian curvature in the domain $D_2$ outwards from the light ray, as opposed to the usual treatment where the deflection is determined by the mass enclosed within the impact parameter (e.g. \cite{schneider} p 231 and \cite{straumann} p 277). Now since the optical Schwarzschild metric is asymptotically Euclidean, we can take $\kappa(\gamma_P)\rmd t/\rmd \phi=1$ on the circular boundary of $D_2$. Moreover, because of the weak deflection limit, we may assume that the light ray is given by $ r(t)=b/\sin \phi $ at zeroth order with impact parameter $b\gg 2\mu$. Hence (\ref{gb3}) together with (\ref{schwarz}) implies
\begin{eqnarray}
\delta&=-\int\int_{D_2} K \rmd S\approx \int_0^\pi \int_{b/\sin \phi}^\infty \frac{2\mu}{r^2}\rmd r \rmd \phi \nonumber \\
&=\frac{4\mu}{b}.
\label{schwarz2}
\end{eqnarray}
This is the well-known formula for the Schwarzschild deflection angle in the weak limit (e.g., \cite{schneider} p 25), as required.

Finally, it should be remarked that the negative Gaussian curvature of the optical metric is a rather general feature of black hole metrics, as is the closed photon orbit forming part of $\partial D_2$. The existence of photon spheres for static, spherically symmetric metrics was first discussed in \cite{atkinson}, and more generally, in terms of an energy condition, in \cite{claudel}. A fuller discussion of the Schwarzschild case, including applications of the Gauss-Bonnet theorem to geodesic triangles, can be found in the forthcoming article \cite{gibbons2}.

\subsection{Plummer model}
The next example is one of the simplest realistic (in the sense of section 3) models for a stellar system, the Plummer sphere \cite{plummer}. It is a polytrope with density distribution (e.g., \cite{binney} p 225)
\begin{equation}
\rho(r)=\rho_0\left[1+\left(\frac{r}{r_0}\right)^2\right]^{-5/2}
\label{plummer1}
\end{equation}
where $\rho_0$ is the central density and $r_0$ a scale radius, so that the mass parameter defined in (\ref{b}) becomes
\begin{equation}
\mu(r)=\mu_\infty\left(\frac{r}{r_0}\right)^3\left[1+\left(\frac{r}{r_0}\right)^2\right]^{-3/2}.
\label{plummer2}
\end{equation}
The corresponding metric is therefore an interior solution with finite $\mu_\infty$ proportional to the total Newtonian mass of this model in the non-relativistic limit where $\mu_\infty/r_0 \ll 1$. We can now compute the Gaussian curvature integrand from (\ref{gauss2}) to first order in $\mu_\infty/r_0$, to obtain
\begin{equation}
K\rmd S=-\frac{2\mu_\infty}{r_0}\frac{u}{(1+u^2)^{3/2}}\left(1-\frac{3}{1+u^2}\right)\rmd u \rmd \phi \ + \ \Or \left(\frac{\mu_\infty^2}{r_0^2}\right), 
\label{plummer3}
\end{equation}
where a dimensionless radius $u\equiv r/r_0$ has been introduced for convenience. Since $\mu$ is finite at large radii, the optical metric of the Plummer sphere approximates the optical metric of the Schwarzschild lens and its negative Gaussian curvature at large radii. Of course, there is no event horizon in this model, so the lens centre $L$ is non-singular, and the domain $D_1$ is simply connected. Then the relevant form of the Gauss-Bonnet theorem (\ref{gb1}) requires that $K>0$ somewhere in $D_1$ for focusing to be possible. Indeed, it can be seen from (\ref{plummer3}) that the leading term of the Gaussian curvature changes sign at $u=\sqrt{2}$, and this is in agreement with the more general conclusion mentioned in section 3.

We can now calculate the deflection angle for the Plummer model by considering the domain $D_2$ as in the previous example. From (\ref{gb3}) and (\ref{plummer3}),
\begin{eqnarray*}
\delta&=-\int\int_{D_2} K \rmd S\approx \int_0^\pi \int_{b/(r_0\sin \phi)}^\infty \frac{2\mu_\infty}{r_0}\left[\frac{u}{(1+u^2)^{3/2}}-\frac{3u}{(1+u^2)^{5/2}}\right]\rmd u \rmd \phi \\
&=\frac{4\mu_\infty}{r_0}\frac{b}{r_0}\left[1+\left(\frac{b}{r_0}\right)^2\right]^{-1}
\end{eqnarray*}
after suitable substitutions. This is the weak deflection angle as expected (e.g., \cite{schneider} p 245), and we recover the Schwarzschild deflection angle (\ref{schwarz2}) in the appropriate limit $b \gg r_0$.

\subsection{Singular isothermal sphere}
The last model discussed here is the singular isothermal sphere. Fully relativistic, isothermal solutions of the Tolman-Oppenheimer-Volkoff equation exist \cite{bisno}, and the density distribution of the singular isothermal sphere is, in fact, similar to the global monopole \cite{barriola}. However, since we are presently concerned with the weak deflection limit, it suffices to consider the non-relativistic, singular isothermal sphere with the following density distribution (see, e.g., \cite{straumann} pp 289--90 and \cite{binney} p 226--8) and mass parameter,
\begin{equation}
\rho(r)=\frac{\sigma^2}{2\pi Gr^2} \ \Rightarrow \  \mu(r)=2\sigma^2 r,
\label{iso1}
\end{equation}
using (\ref{b}), where $\sigma^2 \ll 1$ is again the isotropic velocity dispersion of the stellar fluid. This model is a solution of the Tolman-Oppenheimer-Volkoff equation (\ref{tov}) in the non-relativistic limit, that is, to first order in $\sigma^2$. Then (\ref{gauss2}) shows that the Gaussian curvature vanishes for $r>0$ in this limit, so that the equatorial plane in the optical metric is a cone with a singular vertex at $r=0$. Another way to see this is to isometrically embed the optical metric (\ref{optmetric}) in $\mathbb{R}^3$ with cylindrical coordinates $(z,R,\phi)$ such that
\begin{eqnarray*}
\rmd t^2 &=\exp[2(B-A)]\rmd r^2+\exp(-2A)r^2\rmd \phi^2 \nonumber \\
&= \rmd z(r)^2+\rmd R(r)^2+R(r)^2\rmd \phi^2.
\end{eqnarray*}
Using the equations (\ref{b}), (\ref{a}) and (\ref{iso1}), one can find the coefficients of the spacetime metric (\ref{metric}),
\begin{eqnarray*}
\exp(2A)&=C^{-2} r^{4\sigma^2/(1-4\sigma^2)}, \\
\exp(2B)&=(1-4\sigma^2)^{-1},
\end{eqnarray*}
where $C$ is a non-zero constant. Then the embedding yields
\[
R(r)=C r^{(1-6\sigma^2)/(1-4\sigma^2)},
\]
and the equatorial plane in the optical metric is described by a cone in $\mathbb{R}^3$ of the form
\[
z=\sqrt{8\sigma^2}\frac{\left(1-\frac{9\sigma^2}{2}\right)^{1/2}}{1-6\sigma^2}R
\]
in the non-relativistic limit. This corresponds to a deficit angle of $\Delta\approx 8\pi\sigma^2$.

Now in order to find the deflection angle, we can again apply Gauss-Bonnet in the form of equation (\ref{gb3}) which reads
\begin{equation}
\int_0^{\pi+\delta}\kappa (\gamma_P) \frac{\rmd t}{\rmd \phi} \rmd \phi -\pi=0
\label{iso2}
\end{equation}
since $K=0$ in $D_2$. The Gaussian curvature of the circular perimeter curve $\gamma_P$ can be calculated directly from (\ref{kappa}),
\begin{eqnarray*}
\kappa(\gamma_P)\rmd t&=(1-4\sigma^2)^{1/2}\left(1-r\frac{\rmd A}{\rmd r}\right) \rmd \phi\\
&=\frac{1-6\sigma^2}{(1-4\sigma^2)^{1/2}}\rmd \phi
\end{eqnarray*}
using (\ref{a}). The leading term of the deflection angle is therefore
\[
\delta= 4\pi \sigma^2
\]
from (\ref{iso2}), in agreement with the standard treatment (e.g., \cite{straumann} p 290). This shows that, from this point of view, the constant deflection angle of the singular isothermal sphere comes essentially from the deficit angle of the conical optical metric, similar to cosmic string lensing (e.g., see \cite{gibbons}).

\section{Concluding remarks}
In this article, we have introduced a geometrical approach to gravitational lensing theory different from the spacetime and impulse approximation treatments. By applying the Gauss-Bonnet theorem to the optical metric, whose geodesics are the spatial light rays, we found that the focusing of light rays can be regarded as a topological effect. We also gave a new expression (\ref{gb3}) to calculate the deflection angle by integrating the Gaussian curvature of the optical metric outwards from the light ray, in the weak deflection limit. The lens models considered were given by static, spherically symmetric bodies of a non-relativistic, perfect fluid, and we discussed as examples the Schwarzschild lens, the Plummer sphere and the singular isothermal sphere.

It would therefore be interesting to see whether this approach could also be extended and be fruitfully applied to lenses without spherical symmetry, where images are no longer collinear with the lens centre, or to the relativistic strong deflection limit.

\ack
MCW would like to thank Claude Warnick for useful discussions, and the Science and Technology Facilities Council, UK, for funding.

\end{document}